%% file: main.tex
\documentclass[preprint,12pt,authoryear]{elsarticle}

\usepackage{amsmath, amssymb}
\usepackage{booktabs}
\usepackage{graphicx}
\usepackage{subcaption}
\usepackage{array}
\usepackage{tabularx}
\usepackage{longtable}
\usepackage[table]{xcolor}
\usepackage{hyperref}
\usepackage[capitalize]{cleveref}

\hypersetup{
  colorlinks=true,
  linkcolor=blue,
  citecolor=blue,
  urlcolor=blue
}

\newcommand{\Rrho}{R_\rho}
\newcommand{\Pran}{\mathrm{Pr}}
\newcommand{\rms}{\mathrm{rms}}
\newcommand{\Neff}{N_{\mathrm{eff}}}
\newcommand{\Pb}{P_b}
\newcommand{\QS}{Q_S}
\newcommand{\FS}{F_S}
\definecolor{SWFTableShade}{gray}{0.92}

\journal{Deep-Sea Research Part I}

\begin{document}

\begin{frontmatter}

\title{Interfacial Spectral Memory as a State Variable for Finite-Depth Salt-Finger Exchange}

\author[gt]{Sriram P. Kalathoor\corref{cor1}}
\ead{sriram2@gatech.edu}
\cortext[cor1]{Corresponding author.}
\affiliation[gt]{organization={Daniel Guggenheim School of Aerospace Engineering, Georgia Institute of Technology},
  city={Atlanta},
  state={GA},
  country={United States}}

\begin{abstract}
Thermohaline interfaces in the ocean are often treated through local double-diffusive favorability, yet finite interfaces can also inherit roughness from prior waves, stirring, intrusions, and earlier mixing events. Such inherited geometry can matter because salt fingering does not develop from a flat abstract surface in many geophysical settings. We use controlled three-dimensional direct simulations to test whether the spectral state of a finite rough interface changes the pathway by which salt-finger activity develops between adjacent layers. The density ratio, diffusivity ratio, Prandtl number, interface thickness, roughness amplitude, domain, resolution, and analysis window are held fixed; only the imposed roughness spectrum and, for one pair, the realization are changed. Broad low-mode memory produces the largest cumulative salt exchange and the earliest finite-depth contact. High-annulus memory remains localized and intermediate-scale dominated. Mixed memory produces delayed scale transfer and scalar-rich structure that is robust in integrated exchange and broad-memory measures across a second realization, while local plume timing and probe amplitudes remain realization-sensitive. The simulations therefore support treating interfacial spectral memory as an additional state variable for finite-depth double-diffusive exchange, complementary to local thermodynamic descriptors.
\end{abstract}


\end{frontmatter}

\section{Introduction}

Salt fingering occurs when warm, salty water overlies cooler, fresher water, so that heat diffuses faster than salt and the stratification becomes unstable to narrow vertical exchange. The mechanism appears in the salt-fountain thought experiment of \citet{stommel1956perpetual} and in the thermohaline convection theory of \citet{stern1960salt}. It remains important because it provides a route for vertical salt and heat transport in stratified water columns. Reviews by \citet{turner1974double} and \citet{schmitt1994double} place salt fingering within the broader double-diffusive literature, and \citet{radko2013double} gives a modern theoretical synthesis. Practical descriptions often begin with local stability or profile measures such as density ratio and Turner angle \citep{ruddick1983practical,you2002turner}. Flux measurements, tracer structure, and mixing estimates then connect those local indicators to transport in the ocean interior \citep{gargett2003differential,schmitt2005enhanced}.

Local-gradient measures are necessary; finite interfaces also carry geometric state. Ocean interfaces are not always flat surfaces waiting for an instability to act on them. They may carry finite-amplitude roughness inherited from internal waves, shear, intrusions, earlier double-diffusive events, or nearby turbulent motions. Two interfaces can have the same local double-diffusive favorability and the same mean layer contrast while presenting very different geometric states to the subsequent fingering dynamics. A broad coherent displacement, a short-wave rough interface, and a multi-scale rough interface are not equivalent initial states, even if they share the same density ratio.

The central question is whether that inherited geometric state is dynamically disposable or physically consequential. Local thermodynamic parameters would be sufficient for early exchange if the interface roughness spectrum were erased quickly. In contrast, a spectrum that conditions whether activity remains localized, transfers into broad scales, or reaches finite-depth regions becomes part of the physical state of the interface. For interpreting observed interfaces, this distinction is practical as well as theoretical. Measurements can often detect interface thickness, roughness scale, scalar complexity, or local plume passage more readily than they can reconstruct full three-dimensional transport. The unresolved problem is how those observable interface properties relate to exchange pathways.

Prior work establishes several pieces of this problem. Classical stability studies identify the growth-rate setting for fingering-favorable gradients \citep{turner1964newcase,schmitt1979growth}. Finite-length and secondary instability analyses show that fingers need not remain infinitesimal local modes \citep{holyer1984stability,kunze1987limits}. Collective effects and layer-formation theories then connect fingering motions to larger-scale organization \citep{stern1969collective,radko2003mechanism}. Numerical studies have followed three-dimensional salt fingers from secondary instability to chaotic convection \citep{simeonov2009dns}, quantified small-scale fluxes and large-scale instabilities \citep{traxler2011dynamics}, and connected fingering dynamics to spontaneous layer formation \citep{stellmach2011dynamics}. Observations in stratified water masses show that these processes can matter outside idealized theory \citep{schmitt1981form,schmitt2005enhanced}, while staircase studies describe persistent layered structures in double-diffusive systems \citep{merryfield2000origin,radko2014recipes}. Existing studies leave open how a finite rough interface itself influences the subsequent exchange. We test whether finite-amplitude spectral memory in the interface changes the pathway and outcome of exchange before making a mature staircase or universal flux-law statement.

We isolate one question: for the same local double-diffusive parameters and finite-depth geometry, does the inherited interface spectrum change the exchange pathway? In this controlled system, broad low-mode memory creates a fast, broad connection, high-annulus memory remains localized and intermediate-scale dominated, and mixed memory produces delayed scale transfer with rich scalar structure. A second mixed realization preserves the mixed response in integrated and spectral-memory measures while local timing and probe amplitudes vary. All comparisons are made at fixed density ratio, interface thickness, roughness amplitude, and background state.

\section{Background and Physical Gap}
\label{sec:background}

Classical salt-finger theory identifies the basic instability mechanism and the role of differential diffusion. When the vertical gradients of temperature and salinity have the fingering-favorable sign, heat diffuses across a displaced parcel faster than salt, allowing salinity anomalies to maintain buoyancy differences and drive narrow vertical motions. The density ratio, diffusivity ratio, and Prandtl number summarize this local setting in both classical and multicomponent descriptions \citep{turner1985multicomponent,radko2013double}. Those parameters remain central here; all four integrations use the same values. Holding them fixed isolates whether the finite interface from which the instability develops contains additional information.

Studies of observed water columns have extended the classical theory in several directions. Microstructure measurements connect salt fingering to thermocline mixing and water-mass modification \citep{schmitt1981form,schmitt2005enhanced}. Differential-diffusion reviews and Turner-angle climatologies connect the same physics to tracer structure and profile classification \citep{gargett2003differential,you2002turner}. Arctic staircase observations show that double-diffusive interfaces can persist as measurable oceanic structures \citep{timmermans2008ice,shibley2017arctic}. Permanent staircases in the Tyrrhenian Sea provide a complementary Mediterranean example \citep{durante2019permanent}. Together, these studies establish that salt fingering can move beyond an infinitesimal local instability and reorganize the scalar and velocity fields. It also motivates the practical need to relate measurable quantities, such as gradients, layer thicknesses, rough interfaces, and local plume passage, to exchange consequences.

A large fraction of the quantitative literature is naturally organized around homogeneous or locally idealized gradients. This organization is powerful because it isolates the double-diffusive instability itself. For example, \citet{simeonov2009dns} followed three-dimensional salt fingers from secondary instability to chaotic convection, while \citet{traxler2011dynamics} quantified small-scale fluxes and large-scale fingering instabilities. Staircase-forming simulations by \citet{stellmach2011dynamics} further demonstrated how idealized fingering systems can reorganize into layered states. However, a finite ocean interface is a geometrical object, not merely a pointwise gradient. It has thickness, displacement, roughness, lateral scale content, and a history. Internal waves can corrugate the interface; shear and intrusions can strain it; previous mixing can leave residual roughness; nearby layer deformation can create finite-amplitude departures from a flat state. Such features can be measured or inferred in field contexts, but they are not usually treated as state variables in salt-finger exchange estimates.

The gap lies in the finite interface before mature staircase formation. The simulations prescribe finite-amplitude roughness states and follow how the same salt-fingering system uses them. The relevant physical distinction is between local favorability and pathway selection. Local favorability asks whether the interface can support salt fingering. Pathway selection asks whether the resulting activity remains localized, transfers into broader modes, or reaches finite-depth regions of adjacent layers. The second question requires information about the interface geometry.

Here \emph{spectral memory} means that the distribution of initial interface displacement across horizontal scales affects later exchange-relevant quantities after nonlinear evolution has begun. The memory concept requires neither persistence of every initial phase nor a linear imprint of the imposed displacement on the later flow. A broad-displacement interface can give the instability immediate access to layer-scale organization. A short-wave annular interface can seed localized or intermediate-scale activity without producing a broad connection. A mixed interface can create delayed transfer between those tendencies. The resulting differences are physical outcomes of the initial interface state.

Finite-interface exchange therefore calls for a multi-measure description. A profile or local observation may provide interface thickness, roughness scale, salinity-gradient concentration, plume-passage intermittency, or active-layer depth. Such properties are natural companions to staircase and interface observations \citep{kelley2003diffusive,timmermans2008ice}. None of them alone is a universal flux predictor, but together they describe the state through which double-diffusive exchange is routed. The relevant quantities are exchange, finite-depth reach, spectral transfer, interface geometry, and local measurements as physical descriptors.

\begin{figure}[tbp]
  \centering
  \includegraphics[width=\linewidth]{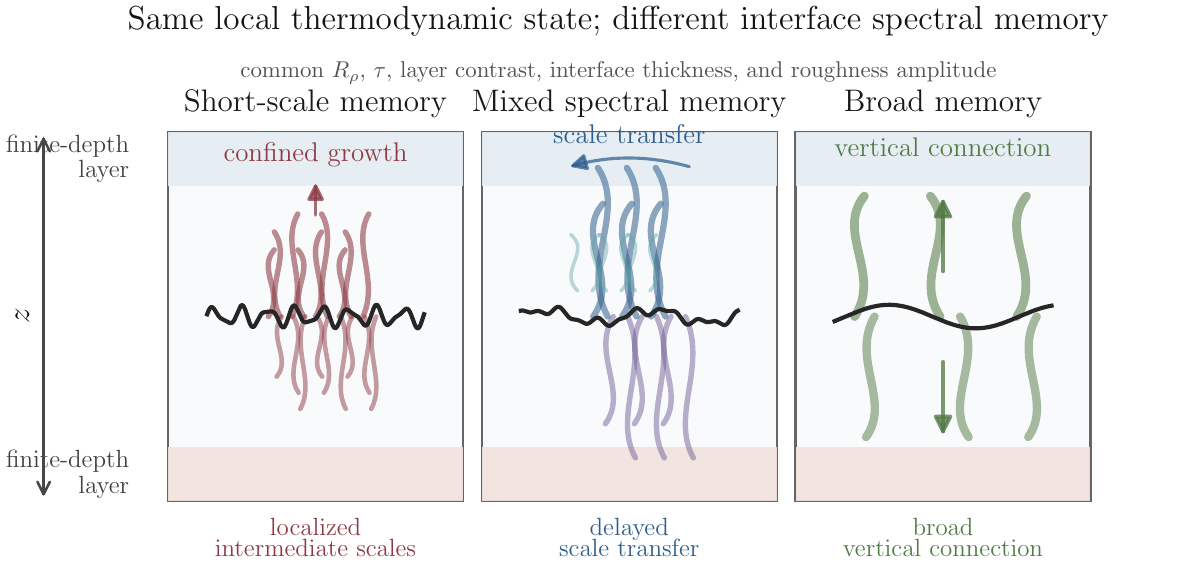}
  \caption{Interface-memory pathways. Interfaces with the same local
  double-diffusive favorability can carry different inherited roughness
  spectra. Broad low-mode memory, short-wave annular memory, and mixed memory
  steer salt-finger activity toward different finite-depth outcomes.}
  \label{fig:concept}
\end{figure}

Inherited interface spectrum changes the subsequent exchange evolution even when local thermodynamic and geometric controls are fixed (\cref{fig:concept}). In the controlled comparison, the local density ratio, diffusivity ratio, interface thickness, and roughness amplitude are held fixed, while the horizontal spectral content of the inherited interface displacement is changed. The three imposed spectra lead to different exchange outcomes under the same local double-diffusive favorability, not separate thermodynamic regimes.

\section{Controlled Interface-Memory Simulations}
\label{sec:setup}

\subsection{Physical comparison}

We consider a finite thermohaline interface in a salt-fingering regime. The density ratio is \(\Rrho=1.2\), the Prandtl number is \(\Pran=7\), and the diffusivity ratio is \(\tau=0.01\). The interface thickness is \(3.0\), and the imposed interface displacement amplitude is \(0.70\). The domain dimensions are \(164.367 \times 82.184 \times 164.367\), resolved on a \(384 \times 192 \times 960\) grid for the four primary comparisons. The analysis window is \(0 \leq t \leq 60\).

The finite-depth domain sets the layer-connection criterion. In an unbounded or vertically periodic model, activity can grow without needing to be classified relative to adjacent finite layers. Here the question is whether activity remains near the interface or approaches the relaxation regions that bound the interior layer. The relaxation regions are not treated as the physics of interest; they define a finite-depth interior over which we can ask whether an exchange becomes layer-connecting during the analyzed window. For that reason, the finite-depth contact measures are reported as distances to the inner relaxation-region edge and as threshold-sensitive first-contact times, not as one binary event.

Only the inherited interface spectrum changes while the thermodynamic and geometric parameters are fixed (\cref{tab:case_inventory}). Broad low-mode memory represents an interface whose displacement is concentrated at large horizontal scales. High-annulus memory represents an interface whose initial displacement is concentrated at shorter annular scales. Mixed-spectrum memory combines broad and short-wave components. A second mixed realization uses the same spectral family with a different realization to test whether the mixed response persists when the phase realization is changed.

The imposed labels identify the initial interface-memory state, not the outcome (\cref{tab:case_inventory}). The low-mode and high-annulus spectra are limiting initial states in the interface-memory space. The mixed spectra test whether combining broad and short-wave memory produces a simple interpolation between those limits or a distinct delayed-transfer response. The second mixed realization tests whether that mixed response survives a changed phase realization.

\begin{table}[tbp]
  \centering
  \caption{Parameters for the interface-memory states. The inherited interface
  spectrum is varied while the local double-diffusive parameters, finite-depth
  domain, grid, roughness amplitude, and analysis window are held fixed.}
  \label{tab:case_inventory}
  \input{tables/case_parameter_inventory.tex}
\end{table}

\subsection{Governing equations and numerical model}

The model solves the three-dimensional incompressible Boussinesq equations for velocity \(\boldsymbol{u}=(u,v,w)\), pressure \(p\), temperature \(T\), and salinity \(S\):
\begin{align}
  \nabla \cdot \boldsymbol{u} &= 0, \\
  \partial_t \boldsymbol{u}
  + \boldsymbol{u}\cdot\nabla \boldsymbol{u}
  &= -\nabla p + \nu \nabla^2 \boldsymbol{u}
  + b\,\hat{\boldsymbol{z}}, \\
  \partial_t T + \boldsymbol{u}\cdot\nabla T
  &= \kappa_T \nabla^2 T, \\
  \partial_t S + \boldsymbol{u}\cdot\nabla S
  &= \kappa_S \nabla^2 S .
\end{align}
The buoyancy \(b\) is computed with a linear equation of state from the temperature and salinity contributions. The nondimensional parameters are \(\Pran=\nu/\kappa_T=7\), \(\tau=\kappa_S/\kappa_T=0.01\), and \(\Rrho=1.2\).

The horizontal directions are periodic. The vertical direction is bounded, with zero-gradient scalar boundary conditions in the two-layer interface formulation and relaxation applied only in far-field sponge regions away from the interface. The equations are integrated with Oceananigans \citep{ramadhan2020oceananigans,silvestri2023oceananigans,wagner2025oceananigans} using a nonhydrostatic pressure projection, centered second-order advection, and scalar diffusivity. The timestep is \(\Delta t=0.0015\), and all four primary integrations are advanced to \(t=60\). Three-dimensional fields and planar slices are saved at the same physical cadence across the comparisons, while point probes sample local plume passage. All comparisons use the common interval \(0\le t\le 60\), with \(t=45\) emphasized as an interior comparison time before broad low-mode activity has spent much of the window inside the far-field relaxation region.

\subsection{Exchange, reach, memory, and local measures}

A small set of measures connects interface memory to finite-depth exchange. The instantaneous salinity exchange is represented by \(\FS(t)\), and cumulative salt exchange is
\begin{equation}
  \QS(0,t) = \int_0^t \FS(t') \, dt' .
  \label{eq:cumulative_exchange}
\end{equation}
The cumulative quantity in \cref{eq:cumulative_exchange} captures differences in peak activity, onset time, and persistence.

Finite-depth reach is measured using salinity and vertical-velocity activity envelopes. For each field, a normalized vertical envelope is compared with the inner edge of the finite-depth relaxation region. The first-contact time is the first time at which the envelope reaches that region for a threshold \(\alpha\). Contact time is not an absolute physical constant; it is a thresholded summary of a continuous activity envelope. Contact ranges over \(\alpha=0.2,0.3,0.4,0.5\), rather than a single threshold, to support localized-versus-connecting language. In practice, the vertical envelope for a field \(q\) is treated as a non-negative activity profile \(A_q(z,t)\), normalized by its instantaneous maximum. The active set is
\begin{equation}
  \mathcal{A}_q(t;\alpha)
  = \{z: A_q(z,t) \geq \alpha \max_z A_q(z,t)\}.
\end{equation}
The active-layer width is the vertical span of \(\mathcal{A}_q\), and the finite-depth distance is the closest distance from this active set to the inner edge of the far-field relaxation region. The definition is intentionally field-based: salinity activity and vertical-velocity activity need not reach the finite-depth region at the same time.

Spectral memory is quantified with radial band fractions. The broad connecting band contains low horizontal wavenumbers associated with broad vertical reach. The intermediate oblique band captures the bridge between broad and short scales. The short-wave annulus contains the initially concentrated shorter roughness band. A fine-scale tail is retained as a resolution and leakage check, but the physical separation comes from the broad, intermediate, and short-wave bands. Band fractions are computed for salinity, vertical velocity, and the salinity-gradient interface height. For a field with horizontal spectral power \(E_q(\boldsymbol{k},t)\), the band fraction is
\begin{equation}
  P_B^q(t) =
  \frac{\sum_{\boldsymbol{k}\in B} E_q(\boldsymbol{k},t)}
       {\sum_{\boldsymbol{k}} E_q(\boldsymbol{k},t)} ,
\end{equation}
where \(B\) is the broad, intermediate, short-wave, or fine-scale radial band. The effective mode count is used as a scalar-richness measure: it is large when power is spread across many horizontal modes and small when power is concentrated in a few modes. The largest exchange response is not necessarily the scalar-richest response.

The interface itself is described from the salinity-gradient surface. We use RMS interface displacement, median gradient thickness, effective interface mode count, and broad interface-height fraction. The quantities are chosen because they correspond to quantities available from profiles or interface observations: interface height, thickness, roughness scale, and scalar complexity. Local probe statistics capture plume passage and local variability at selected locations alongside volume-integrated exchange.

The measure groups are complementary: no single number distinguishes all relevant responses. Cumulative exchange ranks low-mode memory as the strongest exchanger. Effective salinity mode count ranks the mixed spectra as the richest scalar structures. Far-field probe amplitudes emphasize broad plume arrival, while plume-interior covariance can be strong in localized structures if a probe samples an active patch. The interface response is therefore described through exchange, reach, spectral memory, interface state, and local measurements.

\section{Exchange Pathways from Interface Memory}
\label{sec:exchange_pathways}

The inherited spectra produce three distinct responses. Broad low-mode memory forms the fast, broad connection, with the largest cumulative salt exchange, the earliest finite-depth contact, the strongest broad salinity and vertical-velocity memory, and the largest interface displacement and thickening. High-annulus memory is the localized intermediate-scale response: it is dynamically active, but its activity remains concentrated near the interface and does not transfer strongly into broad layer-connecting modes over the analyzed window. Mixed memory forms a delayed transfer response with intermediate cumulative exchange, high scalar richness, delayed broadening, and a signature that repeats across a second realization.

\begin{figure}[tbp]
  \centering
  \includegraphics[width=\linewidth]{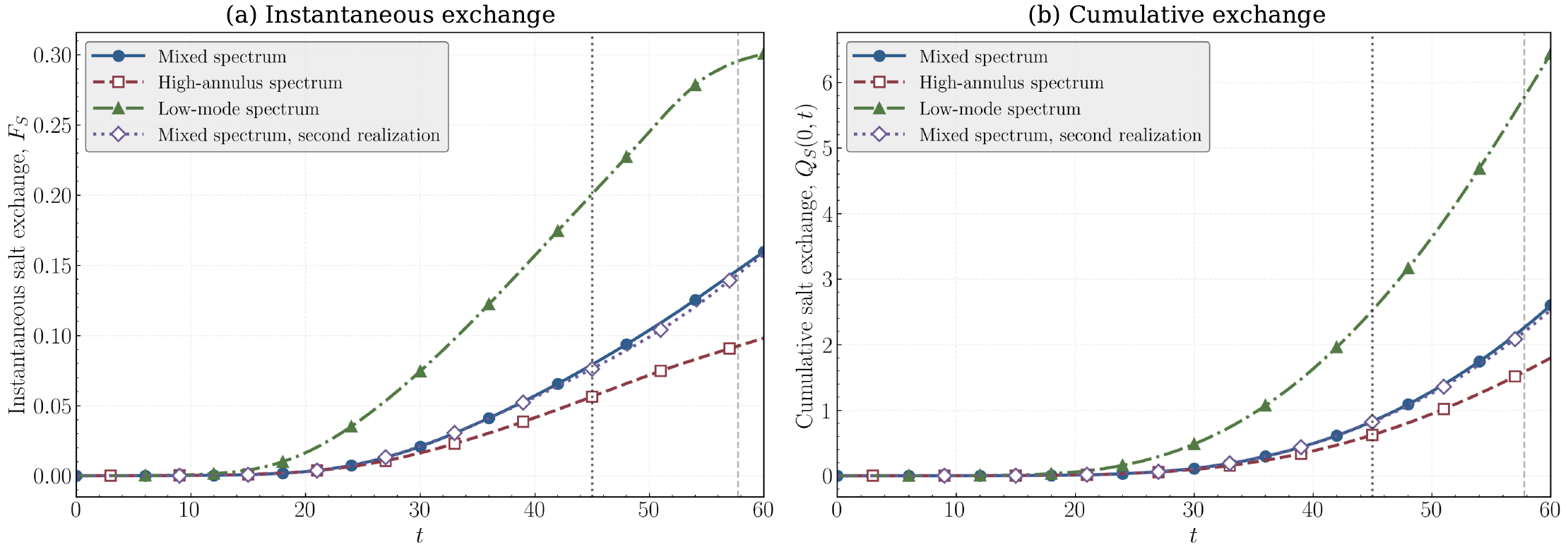}
  \caption{Salt exchange histories. Broad low-mode memory produces the
  strongest cumulative salt exchange, high-annulus memory remains weakest,
  and the two mixed realizations stay close in integrated exchange despite
  differing in local timing and texture.}
  \label{fig:exchange_histories}
\end{figure}

Inherited interface spectrum changes cumulative salt exchange while the local double-diffusive parameters are held fixed (\cref{fig:exchange_histories,tab:state_vector}). At the primary comparison time \(t=45\), the low-mode cumulative exchange is \(\QS(0,45)=2.522\). The corresponding values are \(0.831\) for the mixed spectrum, \(0.821\) for the mixed replicate, and \(0.623\) for high-annulus memory. By this time, broad low-mode memory has exchanged about three times as much salt as the mixed responses and about four times as much as high-annulus memory. The same ordering persists over the full window: \(\QS(0,60)=6.438\) for low-mode memory, \(2.600\) and \(2.530\) for the two mixed realizations, and \(1.797\) for high-annulus memory.

\begin{table}[tbp]
  \centering
  \caption{Exchange, interface, spectral, and local measures. Values combine cumulative exchange,
  finite-depth contact, interface thickness and displacement, scalar richness,
  and broad-band memory.}
  \label{tab:state_vector}
  \input{tables/observation_facing_state_vector_compact.tex}
\end{table}

The exchange separation is more than a shift in timing along one common growth curve. High-annulus memory remains weak throughout the window, while low-mode memory separates early and continues to accumulate exchange. The mixed realizations remain much closer to each other than to either limiting state, which makes the mixed response repeatable rather than a single realization-specific event. The spectral state of the interface conditions the exchange history even though the local fingering parameters are unchanged.

The exchange ranking changes when reach, interface, spectral, and local measures are considered together (\cref{tab:state_vector}). Low-mode memory is the largest exchanger and the earliest finite-depth connector, but it is not the scalar-richest response. High-annulus memory has the weakest cumulative exchange and no finite-depth contact during the analyzed window, but it still has active intermediate-scale structure. Mixed memory occupies the intermediate part of the comparison: their cumulative exchange is close across realizations, their broad-memory measures are similar, and their scalar structure is richer than either endpoint. A single transport number cannot summarize the comparison; exchange, reach, memory, and scalar complexity rank the responses in different but physically interpretable ways.

Heat exchange follows the same primary ordering without identifying a separate pathway. Over \(0 \leq t \leq 60\), cumulative heat exchange is \(2.009\) for low-mode memory, \(0.777\) and \(0.718\) for the two mixed responses, and \(0.252\) for high-annulus memory. The salt-to-heat exchange ratio is larger for high-annulus memory because its heat exchange is especially weak, but this ratio is not used here as a flux-law statement. The transport argument remains centered on salt exchange. Heat exchange provides a supporting consistency check, not an additional state variable in this comparison.

\begin{figure}[tbp]
  \centering
  \includegraphics[width=\linewidth]{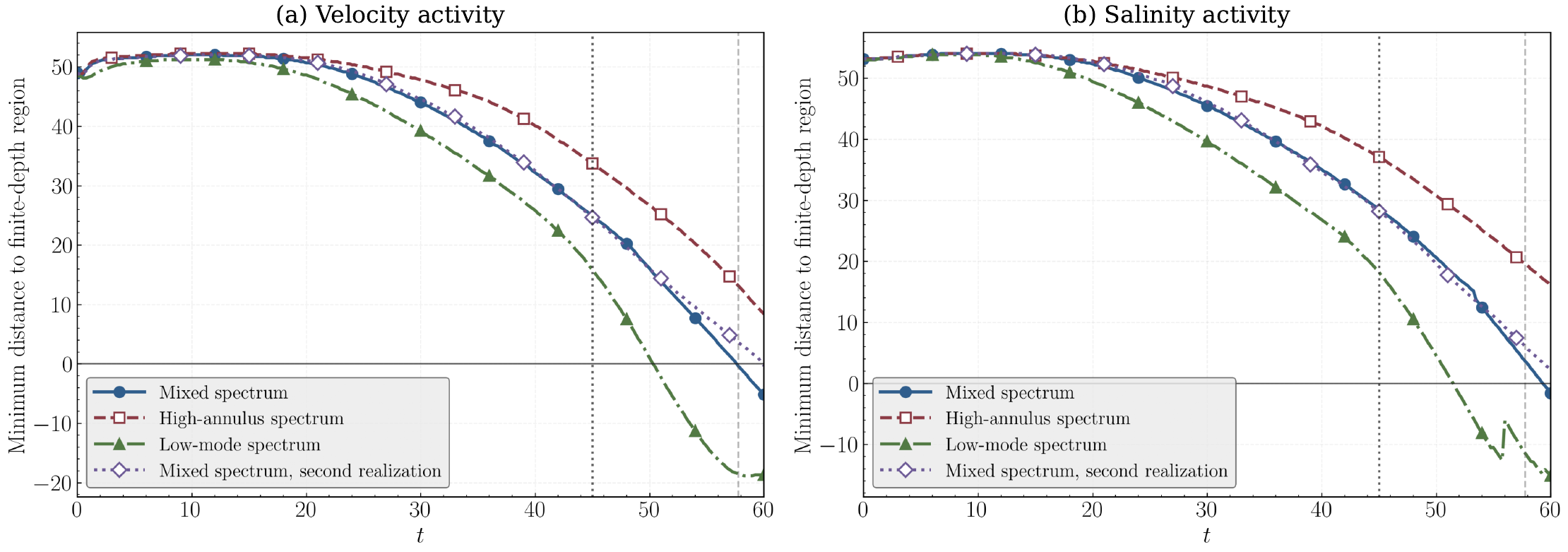}
  \caption{Distance from salinity and vertical-velocity activity envelopes to
  the finite-depth relaxation region. Broad low-mode memory reaches the
  finite-depth region first, high-annulus activity remains separated through
  \(t=60\), and mixed-memory responses approach contact later.}
  \label{fig:finite_depth}
\end{figure}

The finite-depth measures explain why this exchange ordering has layer-scale implications. Low-mode activity extends upward and downward fastest, high-annulus activity remains localized through \(t=60\), and mixed memory approaches finite-depth contact later and with greater threshold sensitivity (\cref{fig:finite_depth,tab:threshold_contact}). Across thresholds, the low-mode vertical-velocity envelope contacts the finite-depth region over \(t=49.75\)--\(51.5\), and the salinity envelope contacts over \(t=51.5\)--\(55.0\). High-annulus memory has no salinity or vertical-velocity contact through \(t=60\) for any tested threshold. The mixed baseline has vertical-velocity contact over \(t=56.25\)--\(59.75\), while salinity contact occurs only at lower thresholds. The mixed replicate has lower-threshold vertical-velocity contact but no salinity contact by \(t=60\). Field dependence matters: vertical motion can extend through the layer before salinity activity fills the same depth range, so velocity contact alone is not treated as complete scalar connection. Conversely, salinity contact at only the lowest thresholds is not promoted to the same status as the low-mode memory, which contacts over the full threshold range in both fields.

\begin{table}[tbp]
  \centering
  \caption{Threshold sensitivity of finite-depth contact times. Times are
  reported for salinity and vertical-velocity envelopes over thresholds
  \(\alpha=0.2,0.3,0.4,0.5\). Values \(>60\) indicate no contact during the
  analyzed window.}
  \label{tab:threshold_contact}
  \input{tables/finite_depth_threshold_contact_compact.tex}
\end{table}

\begin{figure}[tbp]
  \centering
  \includegraphics[width=\linewidth]{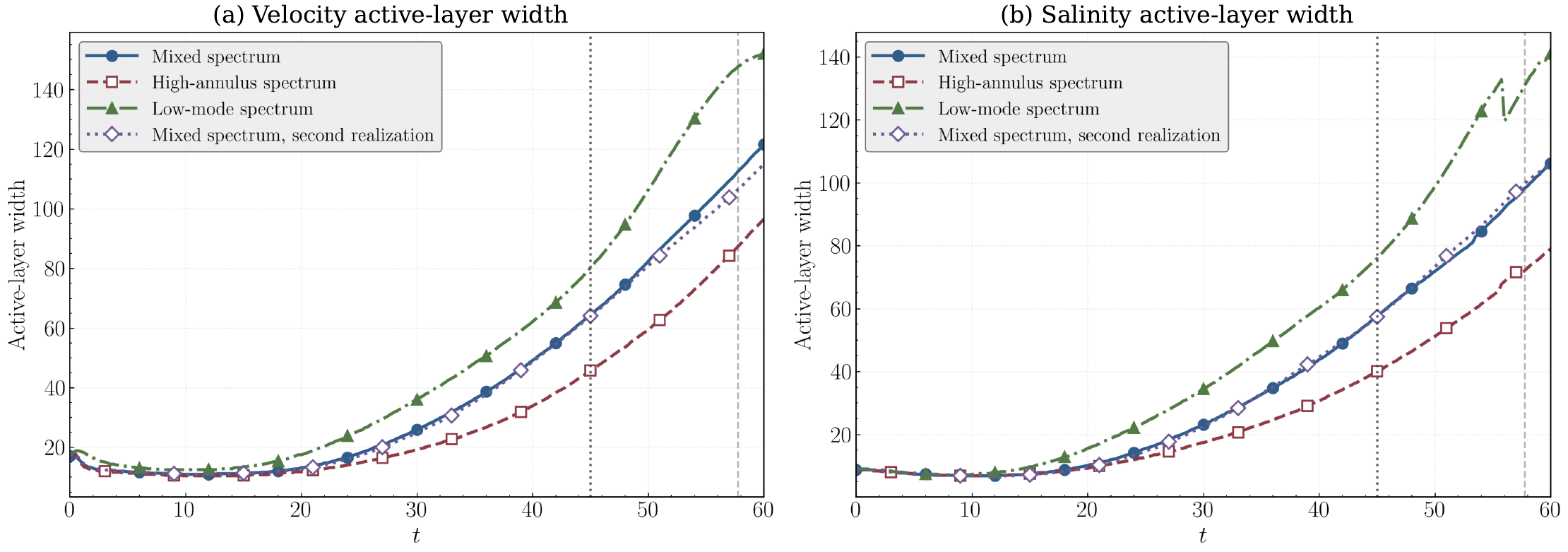}
  \caption{Active-layer width histories for vertical velocity and salinity.
  Low-mode memory broadens fastest, high-annulus memory remains most compact,
  and the mixed responses broaden later while remaining close to each other.}
  \label{fig:active_width}
\end{figure}

Active-layer widths show the same separation before the first low-mode contact. First-contact times identify when an activity envelope first reaches the finite-depth region, while the width histories quantify how the active layer grows before that event. At \(t=45\), low-mode memory spans \(80.13\) in the vertical-velocity activity envelope and \(76.02\) in the salinity envelope. High-annulus memory spans only \(45.71\) and \(40.06\). The mixed baseline and replicate remain close to one another, with \(w\)-activity widths \(64.55\) and \(64.03\), and salinity widths \(57.70\) and \(57.36\). By \(t=57.75\), low-mode activity has already entered the finite-depth relaxation region, high-annulus activity remains separated, and both mixed responses occupy the delayed broadening range (\cref{fig:active_width}). The separation reflects continuous broadening, not only a threshold-crossing event.

The threshold ranges prevent a single cutoff from defining a universal event. Low-mode memory contacts the finite-depth region across all tested thresholds for both salinity and vertical velocity. High-annulus memory contacts none of them. Mixed memory is field-dependent and threshold-sensitive. The finite-depth result reinforces the exchange result: broad inherited memory promotes layer connection, short-wave annular memory remains localized, and mixed memory develops a delayed and less deterministic path toward connection.

The mixed realizations separate visual scalar complexity from bulk exchange. At \(t=45\), the mixed realizations have salinity effective mode counts of \(86.7\) and \(98.3\), far larger than the low-mode value \(5.5\), yet low-mode memory carries the largest cumulative exchange. A scalar-rich interface can be physically active without carrying the maximum exchange. This distinction matters for interpreting observations: complex temperature or salinity structure in a section alone cannot rank layer-scale salt exchange.

\section{Spatial Structure and Scale Transfer}
\label{sec:route_anatomy_transfer}

\begin{figure}[tbp]
  \centering
  \includegraphics[width=\linewidth]{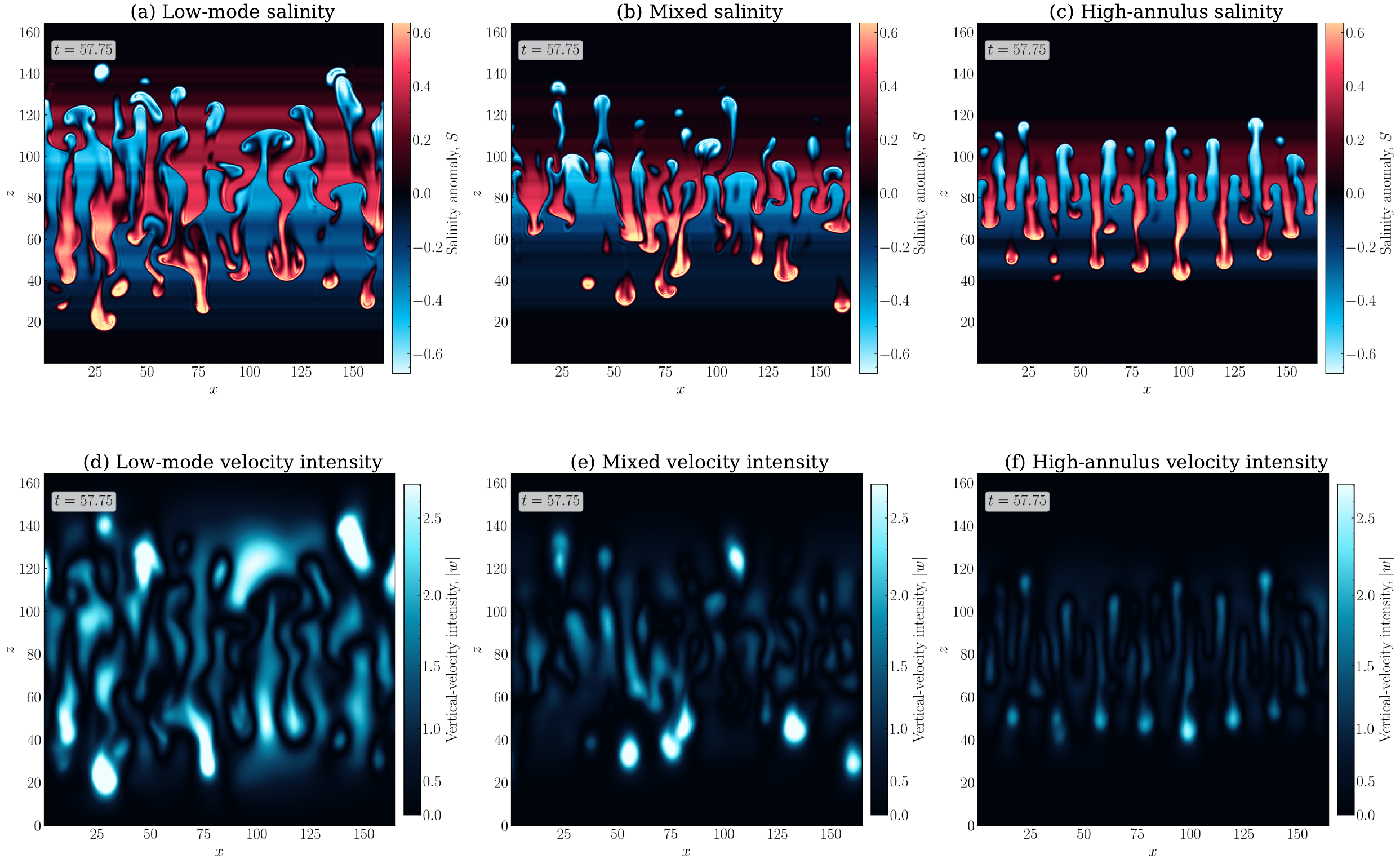}
  \caption{Late-time spatial structure at the comparison time \(t=57.75\). The top
  row gives salinity, and the bottom row gives vertical-velocity intensity.
  Low-mode memory is vertically extensive in both fields, high-annulus memory
  remains more localized, and mixed memory develops delayed multi-scale
  organization.}
  \label{fig:route_anatomy}
\end{figure}

The imposed spectra have direct spatial expression. Low-mode memory forms a broad vertically extensive structure, high-annulus memory remains compact and intermediate-scale dominated, and mixed memory develops multi-scale organization later in the evolution (\cref{fig:route_anatomy}). At \(t=45\), low-mode memory has the largest active vertical reach in the mid-\(y\) salinity field, spanning \(77.90\), and the largest vertical-velocity-intensity reach, spanning \(86.12\). High-annulus memory is more compact, with corresponding spans \(43.66\) and \(47.77\). The mixed baseline and replicate sit between these limits, with salinity spans \(58.56\) and \(53.76\). By \(t=57.75\), these differences intensify: low-mode salinity reach is \(129.10\), high-annulus reach is \(78.93\), and the mixed responses span \(111.98\) and \(105.30\).

The spatial structure explains why the integrated outcomes differ. The low-mode response is broad because active salinity and velocity occupy much of the finite layer depth. High-annulus memory is localized because activity remains closer to the interface and does not approach the finite-depth region. The mixed response is delayed because it develops significant reach only later, and it is scalar-rich because its lateral and interfacial patterns populate more scales than either limiting state. Paired salinity and vertical-velocity-intensity fields are interpreted together because neither field alone carries the complete physical signal. Salinity marks the scalar exchange footprint, including where the interface has been stretched and where scalar anomalies have filled the finite layer. Vertical-velocity intensity marks the active plume pathway that can precede or exceed the scalar footprint. Agreement between the two fields supports layer connection; separation between them signals delayed scalar filling or localized plume passage.

The spectral measures quantify this spatial structure. At \(t=45\), broad salinity power is \(0.795\) for low-mode memory, \(0.405\) for the mixed baseline, \(0.420\) for the mixed replicate, and only \(0.068\) for high-annulus memory (\cref{tab:spectral_memory}). Broad vertical-velocity power at the same time is \(0.948\), \(0.658\), \(0.676\), and \(0.090\). The high-annulus memory is instead dominated by intermediate scales, with intermediate salinity and vertical-velocity fractions \(0.799\) and \(0.874\). High-annulus memory is dynamically organized even though it is weakly exchanging. It remains active, but its organization is concentrated in the intermediate band rather than in the broad connecting band. The mixed responses, by contrast, carry substantial broad-band power by the primary comparison time while preserving scalar richness. These measures separate three physical properties that are easy to conflate: exchange strength, broad vertical connection, and scalar spectral richness.

\begin{table}[tbp]
  \centering
  \caption{Spectral-memory measures. Broad and intermediate band fractions are
  reported for salinity and vertical velocity at \(t=45\), and for the
  salinity-gradient interface height at \(t=57.75\).}
  \label{tab:spectral_memory}
  \input{tables/spectral_memory_compact_table.tex}
\end{table}

\begin{figure}[tbp]
  \centering
  \includegraphics[width=\linewidth]{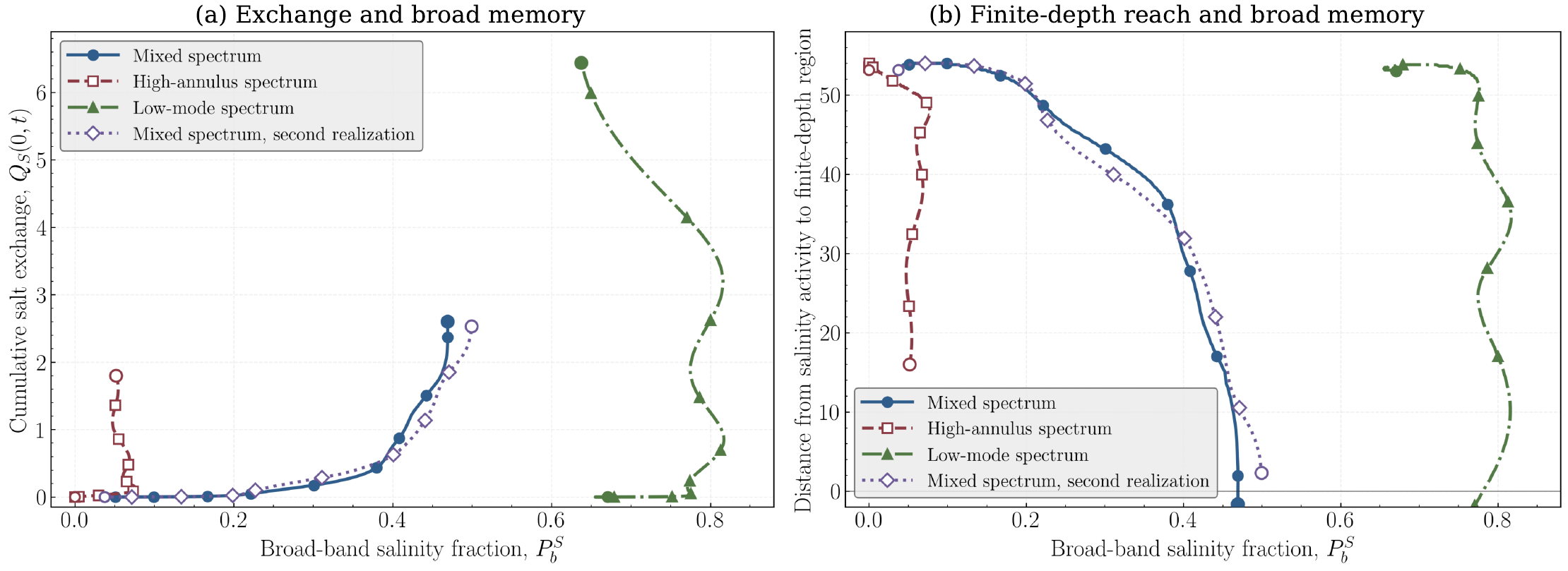}
  \caption{Trajectories of exchange, reach, and broad memory. Each path is
  parametrized by time rather than by a single independent state variable,
  tracing how each interface-memory state moves through exchange and reach
  space.}
  \label{fig:mechanism_state}
\end{figure}

Spectral transfer and cumulative exchange evolve together (\cref{fig:mechanism_state}). Low-mode memory rapidly occupies the broad-memory, high-exchange region. High-annulus memory remains low in broad-band power and cumulative exchange. Mixed memory follows a delayed arc: it begins with substantial shorter-scale content and transfers salinity and velocity power into broad scales over time. The mixed replicate follows the same broad arc even though the local plume phases and contact details differ.

Together, these trajectories explain the scalar-richness result. Mixed memory acts through delayed transfer with a broader population of active scalar scales. High-annulus memory remains active at intermediate scales without efficiently producing broad connection during the analyzed window. Low-mode memory has the initial broad displacement that most directly projects onto the scales that connect the finite-depth layer.

\section{Interface State, Local Measurements, and Robustness}
\label{sec:interface_local_measurements}

\begin{figure}[tbp]
  \centering
  \includegraphics[width=\linewidth]{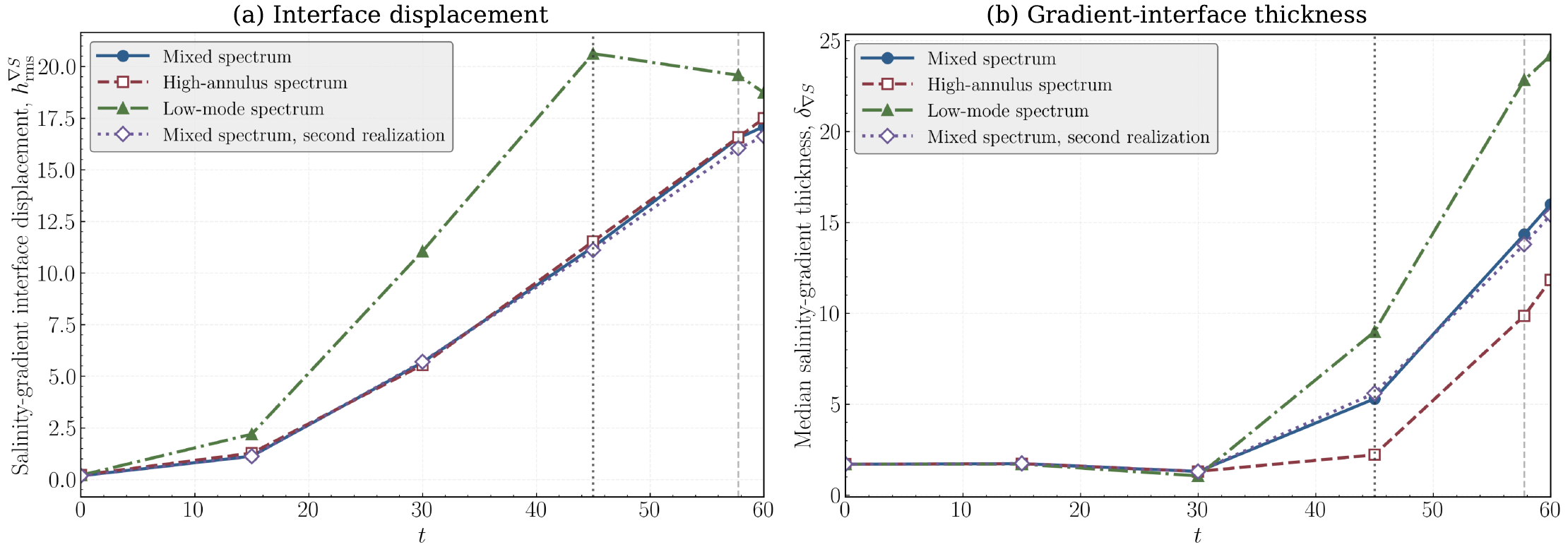}
  \caption{Interface-state histories. The salinity-gradient interface evolves
  differently across memory states in thickness, displacement, and peak
  gradient measures.}
  \label{fig:interface_state}
\end{figure}

The salinity-gradient interface retains the imprint of inherited spectral state after nonlinear growth. At \(t=45\), low-mode memory has RMS interface displacement \(20.62\) and median gradient thickness \(8.99\). High-annulus memory has comparable RMS height to the mixed realizations but a much thinner median gradient thickness \(2.22\). The two mixed realizations have similar displacement and thickness but much larger interface effective-mode counts than the limiting spectra (\cref{tab:interface_state,fig:interface_state}).

\begin{table}[tbp]
  \centering
  \caption{Interface-state coupling. The salinity-gradient interface records
  displacement, thickness, spectral richness, and broad interface-height memory
  that distinguish the exchange responses.}
  \label{tab:interface_state}
  \input{tables/interface_state_coupling_compact.tex}
\end{table}

By \(t=57.75\), broad interface-height fraction is \(0.751\) for low-mode memory, \(0.675\) for the mixed baseline, \(0.680\) for the mixed replicate, and \(0.189\) for high-annulus memory. The salinity-gradient surface retains inherited structure throughout the evolution. Low-mode exchange is tied to large displacement, broad interface memory, and thickening. High-annulus activity remains compact and weakly broad. Mixed memory carries delayed broadening and a scalar-rich interface population. The selected comparison times are part of a coherent interface evolution (\cref{tab:interface_state,fig:interface_state}). Low-mode thickening and displacement grow as activity becomes broad and connecting. High-annulus memory maintains a thinner, less broad interface. The mixed realizations remain close in broad interface-height memory while differing more in local scalar texture. Inherited memory is expressed as an evolving interface state as well as in selected endpoint values.

Several coupled measures describe the interface better than any single scalar predictor. RMS displacement, gradient thickness, effective mode count, and broad interface-height fraction each describe a different property of the same physical interface. A profile that measured only displacement would miss the scalar-rich character of mixed memory; a spectral image that measured only effective mode count would miss the stronger integrated exchange of the low-mode memory. Interface-state measures matter because field observations may recover some aspects of interface geometry more readily than the full three-dimensional exchange field.

Local probe measures provide complementary but more phase-sensitive information (\cref{tab:probes}). They measure plume passage and scalar variability at selected locations, not the volume-integrated exchange. Low-mode memory has the largest far-field vertical-velocity RMS, \(0.908\), and the largest plume-interior local \(|wS'|\), \(0.0492\). High-annulus memory has the smallest far-field \(w_{\rms}\), \(0.123\), consistent with its localized response. However, its plume-interior local \(|wS'|\), \(0.0303\), is not the smallest because localized active structures can pass through selected probe locations.

\begin{table}[tbp]
  \centering
  \caption{Local probe measures. Probe records summarize near-interface
  scalar variability, plume-interior covariance, plume occupancy, and
  far-field plume arrival. They identify local plume passage but do not replace
  integrated exchange.}
  \label{tab:probes}
  \input{tables/probe_intermittency_observation_compact.tex}
\end{table}

The mixed realizations have the same sampling limitation. They are close in cumulative exchange and broad-memory measures, but local probe amplitudes differ. The plume-interior \(|wS'|\) is \(0.0023\) for the mixed baseline and \(0.0199\) for the replicate, while near-interface salinity variability is \(0.106\) and \(0.218\). Far-field event peaks follow the broad-connection picture more clearly, with \(|w|=0.843\) for low-mode memory, \(0.607\) for the mixed baseline, \(0.404\) for the mixed replicate, and \(0.339\) for high-annulus memory.

The second mixed realization separates robust response properties from local plume timing. The cumulative salt exchange changes by only \(-1.2\%\) at \(t=45\) and \(-2.7\%\) at \(t=60\) between the mixed baseline and mixed replicate. Broad salinity power at \(t=45\) changes by \(+3.6\%\), broad vertical-velocity power by \(+2.8\%\), and broad interface-height fraction at \(t=57.75\) by \(+0.8\%\). Together, these measures define the repeated mixed response: delayed transfer into broad salinity, velocity, and interface-height structure with intermediate cumulative exchange. Across the two realizations, the mixed response is robust in integrated exchange and broad-memory transfer, not in every local measurement. Salinity contact is delayed enough that it is not guaranteed by \(t=60\), and local near-interface salinity variability changes strongly because a fixed probe can sample different parts of the plume field in two realizations. For field interpretation, this separation is essential: persistence of mixed memory concerns integrated exchange and memory transfer, whereas local plume records are conditional samples of that evolution (\cref{tab:mixed_route_robustness_extension,tab:mixed_route_local_sensitivity_extension}).

\input{tables/mixed_route_robustness_compact.tex}

The local measures are less repeatable because a different realization moves individual structures relative to the fixed probes. This separation distinguishes a persistent response from exact repetition of individual plume events. In the ocean, individual plume positions would not be expected to repeat, but an interface may still carry a repeatable relation between inherited spectrum, finite-depth reach, and exchange outcome.

\section{Oceanographic Implications}
\label{sec:scope_implications}

The simulations indicate that spectral memory can act as a state variable for finite thermohaline interfaces. Local double-diffusive favorability determines whether salt fingering is possible, but it does not uniquely determine the pathway by which exchange develops when the interface already has finite-amplitude structure. For a finite rough interface, local thermodynamic descriptors and geometric descriptors answer different questions. The local descriptors identify the instability regime. The spectral-memory descriptors help identify whether the resulting activity remains localized, transfers into broad modes, or reaches finite-depth regions of the adjacent layers.

Spectral-memory framing complements flux-law and staircase perspectives. A flux law seeks a compact relation between local stratification and turbulent transport. A staircase study often asks how layers form, merge, or persist. The finite-interface problem sits earlier in the pathway, asking how a finite rough interface steers salt-fingering activity before a mature layered state or a universal transport closure is assumed. The central quantities are cumulative exchange, finite-depth reach, spectral transfer, interface state, and local measurements, not only final transport coefficients.

The differences have three immediate implications for interpreting finite thermohaline interfaces. First, roughness spectra can change cumulative exchange and finite-depth reach at fixed \(\Rrho\), so field estimates based only on local gradients may miss one source of variability. Second, scalar-rich mixed memory need not coincide with maximum bulk exchange, so visual or profile complexity should not be used alone as a transport ranking. Third, localized high-annulus activity can produce local plume covariance without producing layer-connecting exchange, so local measurements need interface-state context to infer layer-scale outcome.

The simulations leave open how commonly each roughness spectrum occurs in the ocean, or how a given ocean interface responds under background shear, variable forcing, or evolving large-scale strain. They also do not provide a universal flux law across density ratio, interface thickness, and roughness amplitude. Within the controlled parameter set studied here, inherited spectral memory changes the exchange pathway even when local parameters are held fixed.

The relevant interface state is more specific than roughness amplitude. All four integrations use the same displacement amplitude, yet their exchange responses separate strongly. Broad low-mode displacement immediately projects onto layer-scale motions, short-wave annular displacement promotes localized and intermediate-scale activity, and mixed memory contains both tendencies and evolves through delayed transfer. The state variable is the distribution of roughness across horizontal scales, not roughness magnitude alone.

\section{Conclusions}
\label{sec:conclusions}

Controlled finite-depth salt-fingering integrations show that inherited interface roughness spectrum can condition the pathway and outcome of exchange while local double-diffusive parameters are held fixed. Broad low-mode memory creates a fast, broad connection, high-annulus memory remains localized and intermediate-scale dominated, and mixed memory forms a delayed scalar-rich response.

The differences are quantitatively large. At \(t=60\), cumulative salt exchange is \(6.438\) for low-mode memory, \(2.600\) and \(2.530\) for the two mixed responses, and \(1.797\) for high-annulus memory. Across tested thresholds, low-mode memory reaches the finite-depth region robustly, high-annulus memory does not contact by \(t=60\), and mixed memory approaches connection late and threshold-sensitively.

The spectral and interface measures explain why those outcomes differ. Low-mode memory remains broad and connecting. High-annulus memory retains intermediate-scale organization without transferring strongly into broad scales. Mixed memory transfers into broad salinity, velocity, and interface height over time, while maintaining high scalar richness. The salinity- gradient interface records the response through displacement, thickness, effective mode count, and broad interface-height memory.

Local probe measures are useful for far-field plume arrival and local scalar variability. They complement integrated exchange because they are local and phase-sensitive, while the repeatable pathway behavior is most robust in cumulative exchange, finite-depth reach, spectral-memory transfer, and interface-state measures.

The mixed replicate demonstrates the relevant kind of robustness. Integrated exchange and broad-memory measures remain close across realizations, while local plume timing and near-interface salinity variability change. The repeatability lies in the persistent mixed response, while individual plume events vary with realization.

The simulations support treating interfacial spectral memory as a physical state variable for finite-depth double-diffusive exchange. In geophysical settings, local thermodynamic favorability should be interpreted together with the finite-amplitude geometry of the interface when assessing whether salt-finger activity is likely to remain localized or develop into layer-connecting exchange.

\section*{CRediT Authorship Contribution Statement}

Sriram P. Kalathoor: Conceptualization, methodology, software, validation, formal analysis, investigation, data curation, visualization, writing -- original draft, and writing -- review and editing.

\section*{Declaration of Competing Interest}

The author declares no known competing financial interests or personal relationships that could have appeared to influence the work reported in this article.

\section*{Funding}

This research received no specific grant from any funding agency in the public, commercial, or not-for-profit sectors.

\section*{Data Availability}

The full raw outputs from the four primary simulations exceed 5 TB because they include three-dimensional fields and high-cadence planar slices, and are not deposited as a complete public archive. The derived tabular data underlying the figures and tables, together with case-setup configurations sufficient to reproduce the integrations in Oceananigans, are available as two versions. Version 1 of the dataset (planar salinity and temperature data) is available on Zenodo \citep{kalathoor2026swfdata1} at \url{https://doi.org/10.5281/zenodo.20756111}, and version 2 of the data (probes and timeseries) is available on Zenodo \citep{kalathoor2026swfdata2} at \url{https://doi.org/10.5281/zenodo.20806266}. Full raw outputs or selected raw subsets can be provided where technically feasible.

\section*{Declaration of Generative AI and AI-Assisted Technologies in the
Writing Process}

During the preparation of this work, the author used Grammarly and Overleaf AI for language editing (spelling and grammar checks). The author reviewed and edited the resulting material and takes full responsibility for the content of this article.

\bibliographystyle{elsarticle-harv}
\bibliography{references}

\end{document}

%% file: tables/case_parameter_inventory.tex
\begingroup
\setlength{\tabcolsep}{3pt}
\renewcommand{\arraystretch}{1.16}
\begin{tabularx}{\linewidth}{
  >{\raggedright\arraybackslash}p{0.17\linewidth}
  >{\raggedright\arraybackslash}p{0.19\linewidth}
  >{\raggedright\arraybackslash}X
  >{\centering\arraybackslash}p{0.12\linewidth}
  >{\raggedright\arraybackslash}p{0.25\linewidth}}
\toprule
Case family & Route role & Interface spectrum & Seed & Shared setup \\
\midrule
\rowcolor{SWFTableShade}
Mixed spectrum & Primary mixed-memory case & Mixed broad and short-wave roughness & 20260604 & \(\Rrho=1.2\); \(\Pran=7\); \(\tau=0.01\); grid \(384\times192\times960\); \(t\leq60\) \\
High-annulus spectrum & Short-scale endpoint & Short-wave annulus only & 20260604 & \(\Rrho=1.2\); \(\Pran=7\); \(\tau=0.01\); grid \(384\times192\times960\); \(t\leq60\) \\
\rowcolor{SWFTableShade}
Low-mode spectrum & Broad-scale endpoint & Broad low modes only & 20260604 & \(\Rrho=1.2\); \(\Pran=7\); \(\tau=0.01\); grid \(384\times192\times960\); \(t\leq60\) \\
Mixed spectrum, second realization & Mixed-route robustness realization & Mixed broad and short-wave roughness & 20260617 & \(\Rrho=1.2\); \(\Pran=7\); \(\tau=0.01\); grid \(384\times192\times960\); \(t\leq60\) \\
\bottomrule
\end{tabularx}
\endgroup

%% file: tables/observation_facing_state_vector_compact.tex
\begingroup
\small
\setlength{\tabcolsep}{1.5pt}
\renewcommand{\arraystretch}{1.15}
\begin{tabularx}{\linewidth}{
  >{\raggedright\arraybackslash}p{0.14\linewidth}
  >{\raggedright\arraybackslash}p{0.18\linewidth}
  *{8}{>{\centering\arraybackslash}X}}
\toprule
Case & Pathway &
\shortstack{\(\QS\)\\\((0,45)\)} &
\shortstack{\(\QS\)\\\((0,60)\)} &
\shortstack{\(t_c^w\)} &
\shortstack{\(\delta_S\)} &
\shortstack{\(h'_{\rm rms}\)} &
\shortstack{\(\Neff^S\)} &
\shortstack{\(\Pb^S\)} &
\shortstack{\(\Pb^w\)} \\
\midrule
\rowcolor{SWFTableShade}
Mixed spectrum & Delayed scale-transfer pathway & 0.831 & 2.600 & 57.75 & 5.298 & 11.258 & 86.7 & 0.405 & 0.658 \\
High-annulus spectrum & Localized intermediate-scale pathway & 0.623 & 1.797 & \(>60\) & 2.221 & 11.532 & 3.3 & 0.068 & 0.090 \\
\rowcolor{SWFTableShade}
Low-mode spectrum & Fast broad connecting pathway & 2.522 & 6.438 & 50.50 & 8.989 & 20.625 & 5.5 & 0.795 & 0.948 \\
Mixed spectrum, second realization & Delayed scale-transfer pathway & 0.821 & 2.530 & $\geq 60$ & 5.602 & 11.097 & 98.3 & 0.420 & 0.676 \\
\bottomrule
\end{tabularx}
\endgroup

%% file: tables/finite_depth_threshold_contact_compact.tex
\begingroup
\small
\setlength{\tabcolsep}{3pt}
\renewcommand{\arraystretch}{1.1}
\begin{tabularx}{\linewidth}{
  >{\raggedright\arraybackslash}p{0.20\linewidth}
  >{\centering\arraybackslash}p{0.06\linewidth}
  *{4}{>{\centering\arraybackslash}X}
  >{\raggedright\arraybackslash}p{0.21\linewidth}}
\toprule
Case family & Field &
\shortstack{\(\alpha\)\\0.2} &
\shortstack{\(\alpha\)\\0.3} &
\shortstack{\(\alpha\)\\0.4} &
\shortstack{\(\alpha\)\\0.5} &
Threshold response \\
\midrule
\rowcolor{SWFTableShade}
Mixed spectrum & w & 56.25 & 57.75 & 58.75 & 59.75 & robust contact \\
Mixed spectrum & S & 58 & 59.5 & $>60$ & $>60$ & threshold-sensitive late contact \\
\rowcolor{SWFTableShade}
High-annulus spectrum & w & $>60$ & $>60$ & $>60$ & $>60$ & no contact through t=60 \\
High-annulus spectrum & S & $>60$ & $>60$ & $>60$ & $>60$ & no contact through t=60 \\
\rowcolor{SWFTableShade}
Low-mode spectrum & w & 49.75 & 50.5 & 51 & 51.5 & robust contact \\
Low-mode spectrum & S & 51.5 & 51.5 & 54.75 & 55 & robust contact \\
\rowcolor{SWFTableShade}
Mixed spectrum, second realization & w & 58.5 & 60 & $>60$ & $>60$ & threshold-sensitive late contact \\
Mixed spectrum, second realization & S & $>60$ & $>60$ & $>60$ & $>60$ & no contact through t=60 \\
\bottomrule
\end{tabularx}
\endgroup

%% file: tables/spectral_memory_compact_table.tex
\begingroup
\small
\setlength{\tabcolsep}{3pt}
\renewcommand{\arraystretch}{1.15}
\begin{tabularx}{\linewidth}{
  >{\raggedright\arraybackslash}p{0.20\linewidth}
  *{6}{>{\centering\arraybackslash}X}}
\toprule
Case &
\shortstack{\(S\) broad\\\(t=45\)} &
\shortstack{\(S\) interm.\\\(t=45\)} &
\shortstack{\(w\) broad\\\(t=45\)} &
\shortstack{\(w\) interm.\\\(t=45\)} &
\shortstack{Interface\\broad\\\(t=57.75\)} &
\shortstack{Interface\\interm.\\\(t=57.75\)} \\
\midrule
\rowcolor{SWFTableShade}
Mixed spectrum & 0.405 & 0.273 & 0.658 & 0.190 & 0.675 & 0.092 \\
High-annulus spectrum & 0.068 & 0.799 & 0.090 & 0.874 & 0.189 & 0.570 \\
\rowcolor{SWFTableShade}
Low-mode spectrum & 0.795 & 0.059 & 0.948 & 0.025 & 0.751 & 0.055 \\
Mixed spectrum, second realization & 0.420 & 0.264 & 0.676 & 0.212 & 0.680 & 0.113 \\
\bottomrule
\end{tabularx}
\endgroup

%% file: tables/interface_state_coupling_compact.tex
\begingroup
\small
\setlength{\tabcolsep}{3pt}
\renewcommand{\arraystretch}{1.14}
\begin{tabularx}{\linewidth}{
  >{\raggedright\arraybackslash}p{0.22\linewidth}
  *{6}{>{\centering\arraybackslash}X}}
\toprule
Case family &
\shortstack{\(\QS\)\\\((0,45)\)} &
\shortstack{\(h_{\mathrm{rms}}^{\nabla S}\)\\\((45)\)} &
\shortstack{\(\delta_{\nabla S}\)\\\((45)\)} &
\shortstack{\(\Neff^{\nabla S}\)\\\((45)\)} &
\shortstack{\(\Pb^{\eta_{\nabla S}}\)\\\((57.75)\)} &
\shortstack{\(\QS\)\\\((0,60)\)} \\
\midrule
\rowcolor{SWFTableShade}
Mixed spectrum & 0.831 & 11.26 & 5.30 & 11.8 & 0.675 & 2.600 \\
High-annulus spectrum & 0.623 & 11.53 & 2.22 & 2.3 & 0.189 & 1.797 \\
\rowcolor{SWFTableShade}
Low-mode spectrum & 2.522 & 20.62 & 8.99 & 2.8 & 0.751 & 6.438 \\
Mixed spectrum, second realization & 0.821 & 11.10 & 5.60 & 12.2 & 0.680 & 2.530 \\
\bottomrule
\end{tabularx}
\endgroup

%% file: tables/probe_intermittency_observation_compact.tex
\begingroup
\small
\setlength{\tabcolsep}{2pt}
\renewcommand{\arraystretch}{1.14}
\begin{tabularx}{\linewidth}{
  >{\raggedright\arraybackslash}p{0.15\linewidth}
  >{\raggedright\arraybackslash}p{0.18\linewidth}
  *{6}{>{\centering\arraybackslash}X}}
\toprule
Case & Pathway &
\shortstack{\(\sigma_S^{near}\)} &
\shortstack{\(\langle |wS'| \rangle\)\\plume} &
\shortstack{\(f\)\\plume} &
\shortstack{\(|w|\)\\plume} &
\shortstack{\(w_{\rms}\)\\far} &
\shortstack{\(|w|\)\\far} \\
\midrule
\rowcolor{SWFTableShade}
Mixed spectrum & Delayed scale-transfer & 0.106 & 0.0023 & 0.387 & 0.761 & 0.331 & 0.607 \\
High-annulus spectrum & Localized intermediate-scale & 0.033 & 0.0303 & 0.282 & 0.477 & 0.123 & 0.339 \\
\rowcolor{SWFTableShade}
Low-mode spectrum & Fast broad connection & 0.007 & 0.0492 & 0.333 & 0.992 & 0.908 & 0.843 \\
Mixed spectrum, second realization & Delayed scale-transfer & 0.218 & 0.0199 & 0.499 & 0.627 & 0.264 & 0.404 \\
\bottomrule
\end{tabularx}
\endgroup

%% file: tables/mixed_route_robustness_compact.tex
\begin{table}[tbp]
\centering
\caption{Large-scale robustness measures for two realizations of the mixed
spectrum.}
\label{tab:mixed_route_robustness_extension}
\footnotesize
\setlength{\tabcolsep}{2pt}
\renewcommand{\arraystretch}{1.08}
\begin{tabularx}{\linewidth}{
  >{\raggedright\arraybackslash}p{0.16\linewidth}
  >{\raggedright\arraybackslash}p{0.28\linewidth}
  >{\centering\arraybackslash}p{0.10\linewidth}
  >{\centering\arraybackslash}p{0.10\linewidth}
  >{\centering\arraybackslash}p{0.10\linewidth}
  >{\raggedright\arraybackslash}X}
\toprule
Group & Quantity & Baseline & Seed & Change & Interpretation \\
\midrule
\rowcolor{SWFTableShade}
Exchange & Cumulative salt exchange, 0--45 & 0.8307 & 0.8209 & -1.2\% & Integrated exchange is nearly unchanged over the primary window. \\
Exchange & Cumulative salt exchange, 0--60 & 2.6 & 2.53 & -2.7\% & Full-window exchange remains close to the baseline mixed response. \\
\rowcolor{SWFTableShade}
Finite-depth reach & Threshold-0.3 velocity contact time & 57.75 & 60 & +3.9\% & Velocity reach is delayed in the replicate but remains a late mixed-route feature. \\
Finite-depth reach & Threshold-0.3 salinity contact time & 59.5 & not observed &  & Salinity contact is timing-sensitive and not guaranteed by \(t=60\). \\
\rowcolor{SWFTableShade}
Spectral memory & Broad salinity fraction at \(t=45\) & 0.4054 & 0.4199 & +3.6\% & Both realizations transfer salinity into broad scales by the primary comparison time. \\
Spectral memory & Broad velocity fraction at \(t=45\) & 0.6576 & 0.6763 & +2.8\% & Velocity broadening is robust across the two mixed realizations. \\
\rowcolor{SWFTableShade}
Interface memory & Broad interface-height fraction at \(t=57.75\) & 0.6746 & 0.6802 & +0.8\% & Late interface-height memory is nearly identical across realizations. \\
\bottomrule
\end{tabularx}
\end{table}

\begin{table}[tbp]
\centering
\caption{Local and spatial sensitivity measures for two realizations of the
mixed spectrum.}
\label{tab:mixed_route_local_sensitivity_extension}
\footnotesize
\setlength{\tabcolsep}{2pt}
\renewcommand{\arraystretch}{1.08}
\begin{tabularx}{\linewidth}{
  >{\raggedright\arraybackslash}p{0.16\linewidth}
  >{\raggedright\arraybackslash}p{0.28\linewidth}
  >{\centering\arraybackslash}p{0.10\linewidth}
  >{\centering\arraybackslash}p{0.10\linewidth}
  >{\centering\arraybackslash}p{0.10\linewidth}
  >{\raggedright\arraybackslash}X}
\toprule
Group & Quantity & Baseline & Seed & Change & Interpretation \\
\midrule
\rowcolor{SWFTableShade}
Interface state & Interface effective mode count at \(t=45\) & 11.78 & 12.17 & +3.3\% & Both realizations retain a scalar-rich interface population. \\
Interface state & Median gradient thickness at \(t=45\) & 5.298 & 5.602 & +5.7\% & Interface thickness is close, with the replicate slightly thicker at \(t=45\). \\
\rowcolor{SWFTableShade}
Spatial anatomy & Mid-y salinity reach at \(t=45\) & 58.56 & 53.76 & -8.2\% & Field anatomy is close but not phase-identical. \\
Spatial anatomy & Mid-y vertical-velocity intensity reach at \(t=57.75\) & 120.5 & 114.9 & -4.7\% & Both realizations show late vertical velocity expansion. \\
\rowcolor{SWFTableShade}
Local probes & Far-field post-window w rms & 0.3311 & 0.2644 & -20.2\% & Far-field local amplitude is realization-sensitive but remains intermediate. \\
Local probes & Near-interface post-window S standard deviation & 0.1057 & 0.2181 & +106.3\% & Local scalar variability is strongly phase/location sensitive. \\
\bottomrule
\end{tabularx}
\end{table}